\documentclass{mem}
\usepackage{natbib}\usepackage{txfonts}\usepackage{balance}
\usepackage{graphicx}
\usepackage[a4paper]{hyperref}
\idline{9999}{1}
\begin{document}
\def\teff{$T\rm_{eff }$}
\def\kms{$\mathrm {km s}^{-1}$}

\title{
WFXT synergies with next generation radio surveys
}

   \subtitle{}

\author{
P. \,Padovani
          }

  \offprints{P. Padovani}

\institute{
European Southern Observatory, Karl-Schwarzschild-Str. 2,
D-85748 Garching bei M\"unchen, Germany
\email{ppadovan@eso.org}
}

\authorrunning{Padovani }

\titlerunning{WFXT synergies with next generation radio surveys}

\abstract{I highlight the synergies of the Wide Field X-ray Telescope
  (WFXT) with the next generation radio surveys, including those to be
  obtained with the Australian Square Kilometre Array Pathfinder and the
  Square Kilometre Array, and discuss the overlap between the X-ray and
  radio source populations. WFXT will benefit greatly from the availability
  of deep radio catalogues with very high astrometric precision, while on
  the other hand WFXT data will be vital for the identification of faint
  radio sources down to $\approx 50~\mu$Jy.

\keywords{galaxies: active --- galaxies: starburst --- radio continuum: 
galaxies --- X-rays: galaxies --- surveys --- telescopes} 
}
\maketitle{}

\section{Introduction}

The Wide Field X-Ray Telescope
(WFXT)\footnote{http://www.eso.org/$\sim$prosati/WFXT/Over\-view.html} is a
medium-class mission designed to be about two orders of magnitude more
sensitive than any previous or planned X-ray mission for large area surveys
and to match in sensitivity the next generation of wide-area optical,
infrared, and radio surveys \citep[see][for details]{gia09,mur09}

In five years of operation, WFXT will carry out three extragalactic surveys:

\begin{itemize}

\item a wide survey, which will cover most of the extragalactic sky ($\sim
  20,000$ deg$^2$) at $\sim 500$ times the sensitivity, and $20$ times
  better angular resolution ($\sim 5$") of the ROSAT All Sky Survey;

\item a medium survey, which will map $\sim 3000$ deg$^2$ to deep Chandra
  or XMM - COSMOS sensitivities;

\item a deep survey, which will probe $\sim 100$ deg$^2$, or $\sim 1000$
  times the area of the Chandra Deep Fields, to the deepest Chandra
  sensitivity.

\end{itemize}

I explore here the possible WFXT synergies with future radio
surveys. Sect. 2 describes the current status of radio surveys, while a
selection of up-coming and future radio projects is described in
Sect. 3. Sect. 4 deals with the source population in deep radio and X-ray
surveys, while the X-ray/radio synergy is discussed in Sect. 5. My
conclusions are summarised in Sect. 6 As this is {\it not} a review of
future radio projects, only basic information on them will be
provided. Readers wanting to know more should consult the relevant
references and World Wide Web pages.

\section{Current radio surveys}

Currently available radio surveys can be divided, as it is the case for
most observational bands, into two main categories: shallow/large area and
deep/small area \citep[see, e.g., Fig. 1 of][]{no09}. The first group
includes, above 0.5 GHz: the NRAO VLA Sky Survey \citep[NVSS;][]{con98},
which covers 82\% of the sky ($\delta > -40^{\circ}$) at 1.4 GHz down to
2.5 mJy, with a 45" resolution; the Faint Images of the Radio Sky at Twenty
centimeters \citep[FIRST;][]{bec95}, covering 22\% of the sky (the North
Galactic Cap) at 1.4 GHz down to 1 mJy, with a 5" resolution; the Sydney
University Molonglo Sky Survey \citep[SUMSS;][]{mau03}, which maps 63\% of
the sky ($\delta < -30^{\circ}$ and $|\rm b_{II}| > 10^{\circ}$) at 843 MHz
down to $\sim 10$ mJy, with a resolution similar to that of the NVSS. The
second category includes a number of Very Large Array (VLA) small area
surveys below 0.1 mJy at a few GHz, reaching a maximum area of $\sim 2$
deg$^2$ \citep[VLA-COSMOS;][]{bon08} and a minimum flux density $\sim 
15~\mu$Jy at 1.4 GHz \citep[SWIRE;][]{ ow08} and $\sim 7.5~\mu$Jy at 8.4 GHz
\citep[SA 13;][]{ fom02}.

\section{Up-coming and future radio surveys}

Radio astronomy is at the verge of a revolution, which will produce large
area surveys reaching flux density limits way below current ones. I
highlight here some of projects, which are being planned.

\subsection{LOw Frequency ARray}\label{lofar}

The LOw Frequency ARray
(LOFAR)\footnote{http://www.astron.nl/radio-observatory/as\-tronomers/lofar-astronomers}
is a new radio telescope designed and built by ASTRON (the Netherlands
Institute for Radio Astronomy) in collaboration with Dutch universities and
other European partners. LOFAR operates in a largely unexplored region of
the electro-magnetic spectrum (from below 20 up to $\sim$ 240 MHz), and
consists of a distributed interferometric array of dipole antenna stations
that permit large areas of the sky to be imaged simultaneously.

LOFAR will carry out large area surveys at 15, 30, 60, 120 and 200 MHz
reaching different flux density limits \citep[see][for details]{mor09}.
For the largest area planned the 120 MHz survey will reach $\approx 0.5$
mJy, which is equivalent to $\approx 0.1$ mJy at 1.4 GHz for a power-law
$\alpha_{\rm r} = 0.7$ ($S \propto \nu^{-\alpha}$). The resolution is
obviously dependent on the longest baseline and on the observing frequency,
and will at best be $\sim 3$" at 240 MHz. LOFAR has started operations in
2010.

LOFAR will open up a whole new region of parameter space at low radio
frequencies. Based on our knowledge of the spectra of the various classes
of radio sources and LOFAR's sensitivity, the large majority of detections
should be radio- and star-forming galaxies, in contrast with X-ray surveys,
which include mostly radio-quiet AGN (see Sect. \ref{deep}). However, the
deeper surveys will reach fainter radio sources and should have a larger
overlap with the type of objects detected in the X-ray band by WFXT.

\subsection{Expanded VLA}

The Expanded VLA (EVLA)\footnote{http://science.nrao.edu/evla} Project will
modernise and extend the existing VLA. When completed in 2012, the EVLA
will provide the following capabilities: observing frequency between 1 and
50 GHz, reaching as low as 1 $\mu$Jy r.m.s. in 6 hours (i.e., between 5 and
20 times better than the VLA), and resolution as good as $\sim 1$" at 1.5
GHz and 0.03" at 45 GHz.  To the best of my knowledge no large area surveys
are being planned at present but some surveys will be obviously carried out
by individual teams.

\subsection{Evolutionary Map of the Universe}

The Australian Square Kilometre Array [SKA] Pathfinder (ASKAP) will produce
wide-deep radio surveys of the sky at 1.4 GHz. The highest ranked ASKAP
continuum project is the Evolutionary Map of the Universe
(EMU)\footnote{http://www.atnf.csiro.au/people/rnorris/emu}.

The primary goal of EMU is to make a deep survey of the entire southern
sky, extending as far north as $\delta = +30^{\circ}$. By reaching a flux
density limit $\approx 50~\mu$Jy EMU will have $\sim 50$ times more
sensitivity than NVSS, whilst covering a similar area (75\% of the sky)
with a five times better angular resolution (10"). EMU will then provide a
similar gain with respect to previous surveys as WFXT in the X-ray band
\citep[see, e.g., Fig. 1 of][]{no09}.

With a likely start of operations in 2013, the EMU catalogue, which will
include around 70 million sources, should be available to the astronomical
community by around 2015.

Besides ASKAP, other radio telescopes currently under construction in the
lead-up to the SKA include the Allen Telescope
Array\footnote{http://ral.berkeley.edu/ata} (ATA),
Apertif\footnote{http://www.astron.nl/general/apertif/apertif}, and
Meerkat\footnote{http://www.ska.ac.za/meerkat}.

\subsection{Square Kilometre Array}

The Square Kilometre Array (SKA)\footnote{http://www.skatelescope.org} will
represent a true revolution in radio astronomy by combining unprecedented
versatility and sensitivity. It will provide an observing window between 70
MHz and 10 GHz reaching flux density limits well into the {\it nanoJy}
regime. Resolution will likely need to be $< 1$" around GHz frequencies to
avoid confusion, with a baseline extending to at least 3,000 km. The field
of view will be large, up to $\sim 200$ deg$^2$ below 0.3 GHz and possibly
reaching $\sim 25$ deg$^2$ at 1.4 GHz. Timeline for completion is 2020,
with first science with 10\% SKA around 2015 - 2016. Location will be in
the southern hemisphere, either Australia or South Africa.

Many surveys are being planned with the SKA, possibly including an "all-sky" 1
$\mu$Jy survey at 1.4 GHz and an HI survey out to redshift $\sim 1.5$,
which should consist of $\sim 10^9$ galaxies.

\section{The deep radio and X-ray skies}\label{deep}

Before discussing the X-ray/radio synergy it is important to have a look
first at the types of sources that are being detected in the two bands, as
current deep radio and X-ray surveys are sampling somewhat different
populations. For example, the X-ray selected sample of \cite{pol07}, with
an X-ray flux limit $f(2-10$ keV)$~> 10^{-14}$ erg cm$^{-2}$ s$^{-1}$
contains $\sim 97\%$ Active Galactic Nuclei (AGN), $\approx 10\%$ of them
radio-loud (as derived from the radio data provided in their
paper). Similarly, the 1 Ms observations of the Chandra Deep Field South
(CDFS) have shown that, amongst the optically brightest sources, 75\% are
AGN and only 22\% are associated with galaxies [\cite{szo04}; see also,
  e.g., \cite{fer08} and references therein].  On the other hand, deep
($S_{\rm 1.4~GHz} \ge 42~\mu$Jy) radio observations of the VLA-CDFS have
identified $\ga 40\%$ AGN (about half of them radio-loud) and $\la 60\%$
star-forming galaxies (SFG) \citep{pad09}. Therefore, while faint X-ray
sources are mostly radio-quiet AGN, deep radio surveys are revealing SFG
and AGN in almost equal numbers, with only about half of the latter, or 
$\approx 1/5$ of the total, being radio-quiet.

This small population overlap is corroborated by the fractions of sources
detected in one band with counterparts in the other one. Of the radio
sources in the VLA-Extended CDFS (ECDFS) sample of \cite{mil08}
($S_{\rm1.4~GHz} \ge 32~\mu$Jy) only $\sim 34\%$ are found in the 2 Ms X-ray
catalogue of \cite{luo08}. And only $\sim 20\%$ of the X-ray sources in 2 Ms
catalogue have a radio counterpart in the VLA-ECDFS survey (Vattakunnel \&
Tozzi, private communication).

It is important to note that X-ray data, including upper limits, play a
very important role in the identification of faint (sub-mJy) radio sources,
as shown by \cite{pad09}. In fact, the radio-to-optical flux density ratio
is not a very good discriminant between SFG and AGN. Radio power fares
somewhat better but is not helpful in separating SFG from radio-quiet
AGN. On the other hand, high X-ray powers ($L_{\rm x} > 10^{42}$ erg/s)
can only be reached by AGN.

\section{The X-ray/radio synergy}

\begin{figure*}[t!]
\resizebox{\hsize}{!}{\includegraphics[clip=true]{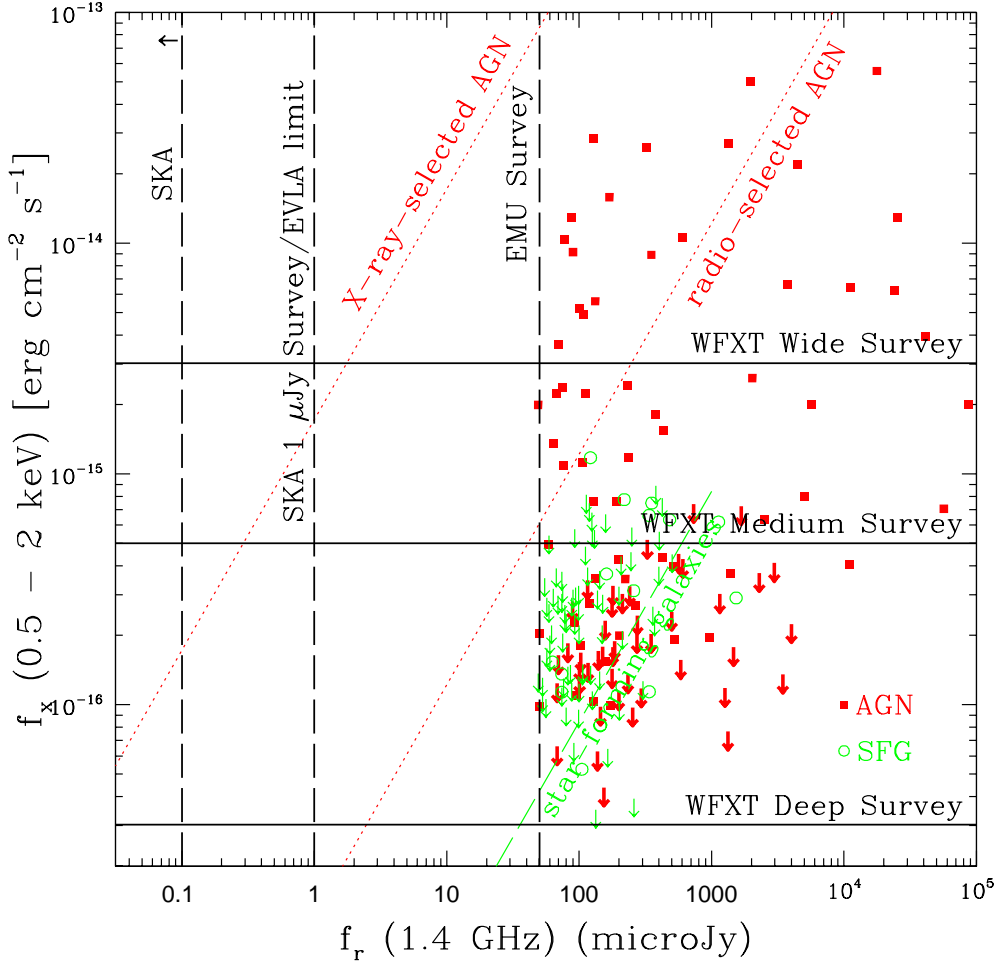}}
\caption{\footnotesize
The 0.5 -- 2 keV X-ray flux vs. 1.4 GHz radio flux density for the AGN
(filled squares) and star-forming galaxies (SFG; empty circles) in the
VLA-CDFS sample \citep{pad09}. Upper limits are also indicated (AGN: thick
lines; SFG: thin lines). The loci of SFG \citep[slanted dashed
  line;][]{ran03}, X-ray selected (mostly radio-quiet) \citep[leftmost
  dotted line, from data in][converted to the 0.5 -- 2 keV band]{pol07} and
radio-quiet, radio-selected AGN \citep[rightmost dotted line from data
  in][]{pad09} are also shown. The position of these loci with respect to
the survey limits determines the fraction of sources of a given class
detected in one band with counterparts in the other. The horizontal solid
lines indicate the limits of the WFXT Wide, Medium, and Deep surveys, while
the two rightmost vertical dashed lines denote the limits of the
Evolutionary Map of the Universe (EMU) and SKA 1 $\mu$Jy surveys, both of
which will cover a large fraction of the sky. The latter represents also
the approximate EVLA r.m.s. level. The leftmost vertical dashed line
at 0.1 $\mu$Jy represents the upper limit for other smaller area SKA
surveys, which will likely be conducted.}
\label{fxfr}
\end{figure*}

Figure \ref{fxfr} plots 0.5 -- 2 keV X-ray flux vs. 1.4 GHz radio flux
density and includes the limits of the WFXT, EMU, and SKA 1 $\mu$Jy surveys
(I am considering here only high-frequency radio surveys for the reasons
discussed in Sect. \ref{lofar}.). Note that, confusion aside, SKA should be
able to detect sources as faint as a few tens of nanoJy.

The loci of SFG, X-ray selected, and radio-quiet, radio-selected AGN are
also shown. These give only order of magnitude estimates as the dispersion
around the mean value can be quite large. For instance, AGN will span the
full range between the two dotted lines in Fig. \ref{fxfr}, with X-ray
(radio) selection favouring sources with high (low) X-ray-to-radio flux
density ratios. The position of these loci with respect to the survey
limits determines the fraction of sources of a given class detected in one
band with counterparts in the other. For example, very few AGN in the WFXT
Wide survey will have a radio counterpart in the EMU survey because the
locus of X-ray selected AGN (leftmost dotted line in Fig. \ref{fxfr}) is to
the left of the EMU limit for $f(0.5 - 2$ keV) $\la 10^{-13}$ erg cm$^{-2}$
s$^{-1}$.

For illustration purposes the fluxes of the AGN and SFG VLA-CDFS sources
are also shown \citep{pad09}. The AGN below the radio-quiet AGN locus are
identified with radio-galaxies.

\subsection{The X-ray survey perspective}

Figure \ref{fxfr} shows that the bulk of the X-ray sources in the WFXT Wide
survey will have a radio counterpart in a possible SKA 1 $\mu$Jy survey. This
should help in the identification work of the 10 million or so expected
objects by also providing very accurate positions. Similarly, most objects
belonging to the Medium survey will be detected in SKA surveys at, or
below, the $\approx 0.3~\mu$Jy level.

Finally, most SFG in the Deep survey will have a radio counterpart already
at the EMU levels, while they will all be detected in an SKA 1 $\mu$Jy
survey. Radio detection of the bulk of the AGN will need much fainter ($<
0.1~\mu$Jy) radio flux limits. This might be accomplished by the SKA 
given also the small area of the WFXT Deep survey ($\sim 100$ deg$^2$). 
All of this, and what follows below (Sect. \ref{radio-pers}), obviously
requires that WFXT surveys are carried out in the southern sky.
 
\subsection{The radio survey perspective}\label{radio-pers}

Figure \ref{fxfr} shows that the bulk of the radio-quiet AGN in the EMU
survey will have an X-ray counterpart in the WFXT Medium Survey, which
should greatly facilitate their optical identification. Overall, one
expects $< 34\%$ X-ray detections, based on the CDFS, which goes deeper in
the X-rays. However, most EMU sources, including SFG, should be detected by
the WFXT Deep survey. The EMU/WFXT combination will provide in this case a
better sample than the VLA-CDFS one on an area $\sim 500$ times larger, for
a total of $> 100,000$ sources.

The bulk of radio-quiet AGN in an SKA 1 $\mu$Jy survey will have an X-ray
counterpart in the WFXT Deep survey. However, at these flux density levels
most objects are expected to be of the SFG type \citep[see,
  e.g.,][]{pad09}, which means that the majority of these radio sources
will not be detected in the X-rays even at the deepest WFXT limit.

Finally, a fraction of the radio-quiet AGN in SKA surveys reaching below
0.1 $\mu$Jy should have an X-ray counterpart in the WFXT Deep survey. Since
at these levels the expected optical magnitudes are very faint even for
unabsorbed sources ($\ga 26$), X-ray information is going to be vital for
source identification.

\section{Conclusions}

Radio astronomy is at the verge of revolutionary advances, which over the
next ten years or so will allow the detection of radio sources as much as
$\ga 100$ times fainter than currently available.

Although at present X-ray and radio surveys detect somewhat different
sources, with AGN making up most of the deep X-ray sky while sharing this
role with star-forming galaxies in the radio band, synergy between the two
bands is already required since, for example, X-ray information is vital to
establish the nature of faint radio sources.

The availability of deep radio catalogues with very accurate source
positions will be a huge asset to WFXT. Similarly, WFXT data will provide
vital help with the identification of faint radio sources down to $\approx
50~\mu$Jy. At lower flux densities the X-ray counterparts of most radio
sources are expected to be fainter than the WFXT deepest limit.

In summary, the combination of future deep radio surveys with WFXT will
shed light on the nature of very faint X-ray and radio sources.
 
\begin{acknowledgements}
I thank Ken Kellermann and Raffaella Morganti for reading the manuscript
and Piero Rosati for useful discussions.
\end{acknowledgements}

\bibliographystyle{aa}

\begin{thebibliography}{}

\bibitem[Becker et al.(1995)]{bec95} Becker, R. H., White, R. L., \&
  Helfand, D. J.\ 1995, \apj, 450, 559

\bibitem[Bondi et al.(2008)]{bon08} Bondi, M.,~et al.\ 2008, \apj, 681, 1129

\bibitem[Condon et al.(1998)]{con98}Condon, J. J.,~et al.\ 1998, \aj, 115, 169

\bibitem[Feruglio et al.(2008)]{fer08} Feruglio, C.,~et al.\ 2008, \aap,
  488, 417

\bibitem[Fomalont et al.(2002)]{fom02} Fomalont, E. et al.\ 2002, \aj, 123,
  2402

\bibitem[Giacconi et al.(2009)]{gia09} Giacconi, R.,~et al.\ 2009, Astro2010: 
The Astronomy and Astrophysics Decadal Survey, Science White Papers (arXiv:0902.4857)

\bibitem[Luo et al.(2008)]{luo08} Luo, B.,~et al.\ 2008, \apjs, 179, 19 

\bibitem[Mauch et al.(2003)]{mau03} Mauch, T.,~et al.\ 2003, \mnras, 343, 1117

\bibitem[Miller et al.(2008)]{mil08} Miller, N. A.,~et al.\ 2008, \apjs,
  179, 114

\bibitem[Morganti et al.(2009)]{mor09} Morganti, R.,~et al. 2009, in
  Panoramic Radio Astronomy: Wide-field 1-2 GHz research on galaxy
  evolution, Gr\"{o}ningen, The Netherlands, June 2009, 
  (arXiv:1001.2384)
  
\bibitem[Murray et al.(2009)]{mur09} Murray, S.,~et al.\ 2009,  Astro2010: 
The Astronomy and Astrophysics Decadal Survey, Science White Papers (arXiv:0903.5272)

\bibitem[Norris et al.(2009)]{no09} Norris, R. P.~et al. 2009, in Panoramic
  Radio Astronomy: Wide-field 1-2 GHz research on galaxy evolution,
  Gr\"{o}ningen, The Netherlands, June 2009, (arXiv:0909.3666)

\bibitem[Owen \& Morrison(2008)]{ow08} Owen, F. N., \& Morrison, G. E.\
  2008, \aj, 136, 1889

\bibitem[Padovani et al.(2009)]{pad09} Padovani, P.,~et al.\ 2009, \apj,
  694, 235

\bibitem[Polletta et al.(2007)]{pol07} Polletta, M.~et al.\ 2007, \apj,
  663, 81

\bibitem[Ranalli et al.(2003)]{ran03} Ranalli, P., Comastri, A., \& Setti,
  G. 2003, \aap, 399, 39

\bibitem[Szokoly et al.(2004)]{szo04} Szokoly, G. P.,~et al.\ 2004, \apjs,
  155, 271

\end{thebibliography}

\end{document}